\documentclass{sig-alternate}
\usepackage{multirow,url}
\usepackage{epstopdf}
\usepackage{graphicx}
\usepackage{rotating}
\begin{document}

\title{Query Representation with Global Consistency on \\ User Click Graph}
\numberofauthors{1}
\author{
\alignauthor
Daqiang Zhang$^1$, Rongbo Zhu$^2$, Shuqiqiu Men$^3$, Vaskar Raychoudhury$^3$\\
$^1$ Department of Computer Science, Nanjing Normal University, China\\
$^2$ Department of Computer Science, South-Central University for Nationalities, China\\
$^3$ Department of Computing, Hong Kong Polytechnic University, China\\
\email{dqzhang@ieee.org}
}

\maketitle

\begin{abstract}
Extensive research has been conducted on query log analysis. A query log is generally represented as a bipartite graph on a query set and a URL set. Most of the traditional methods used the raw click frequency to weigh the link between a query and a URL on the click graph. In order to address the disadvantages of raw click frequency, researchers proposed the entropy-biased model, which incorporates raw click frequency with inverse query frequency of the URL as the weighting scheme for query representation. In this paper, we observe that the inverse query frequency can be considered a global property of the URL on the click graph, which is more informative than raw click frequency, which can be considered a local property of the URL. Based on this insight, we develop the global consistency model for query representation, which utilizes the click frequency and the inverse query frequency of a URL in a consistent manner. Furthermore, we propose a new scheme called inverse URL frequency as an effective way to capture the global property of a URL. Experiments have been conducted on the AOL search engine log data. The result shows that our global consistency model achieved better performance than the current models.
\end{abstract}


\section{Introduction}\label{sec:intro}
Query log analysis has received extensive research attention nowadays, since the exploitation of user feedbacks from query log has been proven to be an
effective and non-intrusive method to improve search quality.
Search engine has been recording user click through information all the time, which can be represented as a bipartite graph. The bipartite graph refers to query and URL in most cases. An edge connects a query and a URL and the edge value generally corresponds to the click frequency.

Many query log analysis models are based on the click graph, in that a certain URL has been clicked by different queries (issued by users) and hence provides the information about the relevance of URL, query and user. A query can be represented as a vector in which each dimension corresponds to the edge value between the query and a URL. Traditional models make use of the raw click frequency (the number of clicks or users) between a query and a URL, which suffers from two problems: first, raw click frequency does not favor unpopular queries or URLs; moreover, ranking models based on raw click frequency often favor already frequently clicked URLs because of inherent bias of clicks~\cite{Craswell:Position-bias}. Therefore it is worth to research on how to improve the representation of click graph before developing any analysis method. To leverage the influence of highly clicked URLs, an entropy-biased model in~\cite{Deng:Entropy-biased} has been proposed to address the disadvantages of raw click frequency by weighting the raw click frequency with {\em inverse query frequency} (IQF), under the assumption that less clicked URLs are more relevant to a given query than heavily clicked ones. The inverse query frequency is inspired by {\em inverse document frequency} (IDF) in text retrieval, and~\cite{Deng:Entropy-biased} incorporates inverse query frequency into user frequency in the same manner as TF-IDF does. Although there are many interpretations~\cite{Robertson:Understanding,Vries:Loss-entropy,Roelleke:Uncovered} about why TF-IDF has proved to be extraordinary robust in text retrieval, utilizing user click frequency in the same manner as TF-IDF may not be appropriate in the context of user click graph.

Different from content-aware text retrieval, the click through information is the implicit feedback from users in that each click denotes a potential association between a query and a URL. A click tends to be more informative than the case in text mining, since a document could contain much irrelevant information. Therefore, our observation is that user frequency and inverse query frequency should be treated differently during query representation in the context of click graph. Consistent with the assumption that less clicked URLs tend to be more relevant to a given query, the inverse query frequency should be more informative than user frequency according to our observation. Moreover, if inverse query frequency can be considered as a global property of each URL on click graph, it is intuitive to develop the global consistency model for query representation, which utilizes user frequency and the global weight of URL on user click graph in a consistent way to achieve better performance, as described in this paper.

The contribution of this paper lies in: 1) we observe that the global nature of the URL plays a central role for query representation on user click graph; 2) A new scheme called {\em inverse URL frequency} (IUF) is presented to specify the global weight of each URL on click graph, and result shows that IUF is superior to IQF in the context of global consistency on click graph; 3) we define the rules for achieving global consistency on click graph, and develop the framework of global consistency model for query representation.

The rest of this paper is organized as follows: the related works are introduced in section~\ref{sec:rl}, and we illustrate global consistency on click graph in section~\ref{sec:gcongraph}. Various query representation models are presented in section~\ref{sec:queryrepresent}, while section~\ref{sec:datacollection} and section~\ref{sec:exp} presents the experimental analysis about the performance of different models under query similarity. Conclusion is made in section~\ref{sec:conclusion}.

\section{Related Work}\label{sec:rl}
Extensive research has been conducted on click graphs to exploit implicit feedback~\cite{Agichtein:Improving}. Frequently studied topics include agglomerative clustering~\cite{Beeferman:Agglomerative}, query clustering for URL recommendation~\cite{Wen:Clustering}, query suggestion~\cite{Mei:Hitting}, which used hitting time to generate semantic consistent suggestions, and rare query suggestion~\cite{Song:Rare-Query}. Moreover,~\cite{Li:Query-intent} worked on query classification through increasing the amount of training data by semi-supervised learning on the click graph instead of enriching feature representation. While there are works studying different aspects of user click information,~\cite{Craswell:Position-bias} revealed that the click probability of a webpage is influenced by its position on the result page. The sequential nature of user clicks has been considered in~\cite{Ji:Global}, whereas \cite{Song:Rare-Query} combined both the click and skip information from users. In addition, having noticed that click graphs are very sparse and the click frequency follows the power law,~\cite{Xue:Optimizing} made use of co-click information for document annotation. Random walk has been applied to click graphs~\cite{Craswell:Random-walk} to improve the performance of image retrieval.~\cite{Gao:Smoothing} also employed random walk to smooth the click graph to tackle the sparseness issue.

In contrast, less work has been carried out on the study of query representation on click graphs.~\cite{Baeza-Yates:Semantic} represented each query as a point in a high dimensional space, with each dimension corresponding to a distinct URL.~\cite{Poblete:Query-sets} introduced the query-set based model for document representation using query terms as features for summarizing the clicked webpages. The entropy-biased model for query representation has been proposed~\cite{Deng:Entropy-biased} to replace raw click frequency on the click graph. It assumed that less clicked URLs are more effective in representing a given query than heavily clicked ones. Thus, the raw click frequency was weighed by the inverse query frequency of the URL. However, the entropy-biased model utilized raw click frequency and inverse query frequency in the same manner as TF-IDF does, which may not be appropriate in the context of click graph. This is because user click information is content-ignorant while text retrieval is content-aware. Our work is closely related to~\cite{Deng:Entropy-biased}, while our contribution is to study how to combine the raw click frequency and the global weight of URL in a consistent way for query representation.

\section{Global Consistency}\label{sec:gcongraph}
\subsection{Preliminaries}
The user click graph is generally regarded as a bipartite graph. In this paper, we consider a bipartite graph evolves query and URL: G = (Q $\cup$ D, E), where the query set Q and document (URL) set D are connected by edges in E. Suppose there are M queries and N documents in total, the bipartite graph can be represented as a rank M$\times$N matrix C, with the entry (i, j) as the edge value of ($q_i$, $d_j$). In most cases, the edge value corresponds to the raw click frequency $c_{ij}$ between query $q_i$ and document (URL) $d_j$, which is the number of times the users click on $d_j$ when $d_j$ is presented to the users as a result for query $q_i$. Thus, a query $q_i$ can be represented as a row vector of C, and a document (URL) $d_i$ corresponds to a column vector of C.

\begin{table}
\centering
\caption{Preliminaries and Notations}
\renewcommand{\arraystretch}{1.2}
{\small
\begin{tabular}{|c|l|} \hline
|Q|&Total number of queries\\ \hline
|U|&Total number of URLs (documents)\\ \hline
\multirow{2}{*}{$q(d_j)$}&Number of distinct queries associated with
\\&URL j\\ \hline
\multirow{2}{*}{$u(d_j)$}&Number of distinct URLs indirectly associated
\\&with URL j via queries\\ \hline
$p(d_j|q_i)$&Transition probability from query i to URL j\\ \hline
$p(q_i|d_j)$&Transition probability from URL j to query i\\ \hline
$p(q_j|q_i)$&Transition probability from query i to query j\\ \hline
$P_{q2d}$&Transition probability matrix for query-URL\\ \hline
$P_{d2q}$&Transition probability matrix for URL-query\\ \hline
$c_{ij}$&Click frequency from query i to URL j\\ \hline
$uf_{ij}$&User frequency from query i to URL j\\ \hline
$v_{ij}$&Edge value obtained from various models\\ \hline
$g(d_j)$&Global weight of URL j on click graph\\ \hline
\end{tabular}
}
\label{tab:preliminaries}
\end{table}

To simplify our illustration, we use $v_{ij}$ to denote the edge value when we apply different models to represent the click graph. For example, $v_{ij} = c_{ij}$ if click frequency is used for graph representation. Note the difference between click frequency $c_{ij}$ and user frequency $uf_{ij}$: $c_{ij}$ corresponds to the raw click frequency regardless of user identities, while $uf_{ij}$ corresponds to number of distinct users who issued query $q_i$ and clicked URL $d_j$. In this work we only consider user frequency since previous work suggested that it is more robust to spurious clicks. The preliminaries and notations used in this paper are listed in Table~\ref{tab:preliminaries}.

\subsection{Global Nature on Click Graph}\label{sec:global-nature}

\begin{figure}
\begin{center}
\resizebox{80mm}{!}{\includegraphics{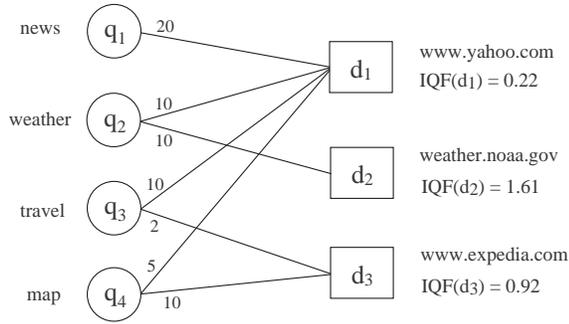}}
\caption{A bipartite graph example.}
\label{fig:fig1}
\end{center}
\end{figure}

A bipartite graph is generally used to represent user click through information, as illustrated in Figure~\ref{fig:fig1}. Traditionally, a query can be represented by the edge values associated with different URLs, such as click frequency ($v_{ij} = c_{ij}$) or user frequency ($v_{ij} = uf_{ij}$). Previous work argued that different query-URL pairs should be treated differently~\cite{Deng:Entropy-biased}. For example, the edge `weather - www.yahoo.com' and the edge `weather - weather.noaa.gov' have the same user frequency ($uf_{21} = uf_{22} = 10$) in Figure 1, but `weather.noaa.gov' should be more relevant to the query `weather' intuitively. To alleviate this problem, IQF ($IQF(d_j) = log(|Q|/q(d_j)$, inverse query frequency)~\cite{Deng:Entropy-biased} has been proposed to improve the weighting scheme. In particular, `weather.noaa.gov', which is associated with one distinct query obtains a larger IQF value than `www.yahoo.com' which associated with four queries ($IQF(d_2) > IQF(d_1)$). Thus after the raw click frequency is multiplied by the IQF value of each URL, the edge `weather - weather.noaa.gov' obtains a larger value than the edge `weather - www.yahoo.com' ($v_{21} < v_{22}$, $v_{ij} = uf_{ij} \times IQF(d_j)$, details can refer to Table~\ref{tab:query-represent}).

With the observation that less clicked URLs are more useful than heavily clicked ones for query representation, the inverse query frequency is inspired by the inverse document frequency (IDF) weighting scheme in text retrieval.
The main different between query representation on click graphs and term
weighting in text retrieval is that on a click graph
clicks are performed by users after the users had acquired some ideas
about the relevance of the documents with respect to their queries based
on the snippets of the results. On the other hand, in text retrieval
some terms (e.g., stop words or common terms in a particular domain
such as ``data'' in a scientific collection) may appear in many documents
and bear no relevance to the topics of the documents. Thus, IDF is incorporated as a term discriminator~\cite{Roelleke:Uncovered} whereas term frequency
remains an important component of the weighting formula since
it indicates the local importance of a term in a document. On a click graph, inverse query frequency (IQF) corresponds to the number of distinct queries associated with each URL, and it contains less noise compared to terms in documents. Instead, click graph suffers from the sparseness problem. In other words, with respect to a certain query, a high IQF value tends to indicate a specific URL while a low IQF value denotes an ambiguous one. Thus, it is inappropriate to incorporate the user frequency into inverse query frequency in the same manner as TF-IDF does.

From the discussion above, inverse query frequency is more informative than user frequency in a click graph. Some early works like PageRank~\cite{Brin:Anatomy} and~\cite{Kleinberg:Authoritative} tried to identify the global properties of documents with link analysis on the hyperlink graph. This motivates us to consider the inverse query frequency as a global property of URL, and to develop the global consistency model for query representation on a click graph, in which the global nature of URLs plays a central role and user frequency should be incorporated in tune with the global nature of the URL. For example, in Figure 1, in the previous entropy-biased model, although `www.expedia.com' has a larger IQF value than `www.yahoo.com', the edge `travel - www.yahoo.com' still larger than that of `travel - www.expedia.com' after the UF-IQF weighting because of the user frequency factor. Instead, for representation with global consistency, the difference lies in: 1)`travel - www.yahoo.com' will obtain a smaller value than `travel - www.expedia.com', consistent with their IQF value of `www.expedia.com' and `www.yahoo.com' respectively; 2)compared with the original graph, the representations of query `travel' and query `map' become more similar since they connect to exactly the same URLs and their edge values will be consistent with the IQF values of the URLs.

The rationale behind this global consistency representation in the context of user click graph is that for a certain query, its contribution to a URL with higher global weight is relatively more than it contributes to a URL with lower global weight, given that less clicked URLs are more globally important than highly clicked ones. And if all the queries are represented in the manner that edge value consistent to the global weight of URL, it is expected to improve the similarity between different queries. The rest of this section illustrates how to specify the global nature of the URL on user click graph, and section~\ref{sec:global-consistency} define the rules for query representation which is consistent with the global nature of URLs.

{\bf Inverse Query Frequency as Global Nature:} Under the assumption that less clicked URLs are more relevant to the given query than heavily clicked ones, {\em inverse query frequency} (IQF) as mentioned before can be employed to denote the global weight of each URL on user click graph, which considers the number of distinct queries associated with a URL:
\begin{equation}g(d_j) = IQF(d_j) = \log(\frac{|Q|}{q(d_j)})\end{equation}
The intuition behind IQF is that less clicked URLs tend to be less ambiguous and thus more important, and the click frequency of a certain URL is proportional to the number of distinct queries it associates.

{\bf Inverse URL Frequency as Global Nature:} Consistent with inverse query frequency, we can further consider the global nature of a URL as the relation between the total number of URLs and the number of distinct URLs ($u(d_j)$) it associates via the queries, which is named as inverse URL frequency (IUF). Take Figure 1 for example, $u(d_2) = 2$ since `weather.noaa.gov' indirectly associates 2 URLs via query `weather'. Like one hop on click graph, IUF further extends the queries associated with a certain URL to the URLs that the queries are connecting:
\begin{equation}g(d_j) = IUF(d_j) = \log(\frac{|U|}{u(d_j)})\end{equation}
The motivation behind IUF is to further consider the similarities among the queries in the case of IQF. If the queries associated with a certain URL are similar, then the number of distinct URLs they are connecting should be less than the number of the queries; on the other hand, if the queries associated with a certain URL are dissimilar, then the number of distinct URLs they are connecting should be no less than the number of queries. For example, consider the following two URLs: `www.yahoo.com' and `www.microsoft.com'. Suppose the numbers of distinct queries associated with the two URLs are the same, which means the two URLs have the same global weight in terms of inverse query frequency (IQF). However, what we can imagine is that the queries associated with `www.yahoo.com' tend to be diverse topics and connect to a larger number of distinct URLs, while the queries associated with `www.microsoft.com' tend to be similar and connect to a smaller number of distinct URLs. In other words, the queries from `www.microsoft.com' are less diverse than the ones from `www.yahoo.com' in terms of the number of distinct URLs connecting them. Therefore, the global weight of `www.microsoft.com' should be larger than `www.yahoo.com' under the assumption that less ambiguous URL is more relevant to a given query, which can be denoted using inverse URL frequency (IUF).

\subsection{Global Consistency Representation}\label{sec:global-consistency}

\begin{figure}
\begin{center}
\resizebox{50mm}{!}{\includegraphics{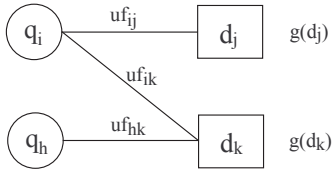}}
\caption{Global consistency illustration.}
\label{fig:fig2}
\end{center}
\end{figure}

We define three rules about what is query representation with global consistency on click graph, illustrated using Figure~\ref{fig:fig2}.

\begin{enumerate}
\item Given a query $q_i$, if $g(d_j) > g(d_k)$, it is essential that $v_{ij} > v_{ik}$ (global consistent) regardless the value of $uf_{ij}$ and $uf_{ik}$.
\label{item:rule1}
\item With respect to a certain query $q_i$, if $g(d_j) = g(d_k)$, and $uf_{ij} > uf_{ik}$, we should achieve $v_{ij} > v_{ik}$ (distinguish between different user frequencies if the URLs they are connecting hold the same global weight).
\label{item:rule2}
\item As for two queries $q_i$ and $q_h$, if $\sum_{d\in D}uf_{id} > \sum_{d\in D}uf_{hd}$ and $uf_{ik} = uf_{hk}$, it is necessary that $v_{ik} < v_{hk}$ (a less diverse query is more relevant to the user intention).
\label{item:rule3}
\end{enumerate}

The above rules play an important role about how to develop query representation model with global consistency, which is presented in section~\ref{sec:queryrepresent}.

\section{Query Representation}\label{sec:queryrepresent}

\subsection{User Frequency Model}
Since previous work~\cite{Deng:Entropy-biased} has demonstrated that user frequency is better than click frequency in eliminating spurious clicks, while our focus is on the global nature of the click graph, we only consider user frequency in this work:
\begin{equation}v_{ij} = uf_{ij}\end{equation}
The edge value between query $q_i$ and URL $d_j$ is represented as the number of distinct users who have issued query $q_i$ and clicked URL $d_j$.

\subsection{Entropy-biased Model}
Inspired by TF-IDF, the entropy-biased model from~\cite{Deng:Entropy-biased} proposed {\em inverse query frequency} for the URL, which is closely related to the discriminative ability of the URL, and URL with smaller IQF value tends to be less discriminative. Therefore it incorporates inverse query frequency into user click frequency to formalize the entropy-biased query representation (UF-IQF):
\begin{equation}v_{ij} = uf_{ij} \times log(\frac{|Q|}{q(d_j)})\end{equation}
The entropy-biased model has the benefit of reducing the influence of heavily clicked URLs and boosting the URLs that have only a few clicks. But what entropy-biased model had ignored is the relative weight between user frequency and inverse query frequency on click graph, which will be addressed in global consistency model.

\begin{table}
\centering
\caption{Query representation for Figure~\ref{fig:fig1} with different models
(Here suppose |Q|= 5 and |U|= 4 to avoid log(1) problem, since in Figure~\ref{fig:fig1} |Q|= 4 and |U|= 3)}
\begin{tabular}{|c|c|c|c||c|c|c|c|} \hline
\multicolumn{4}{|c|}{UF}&\multicolumn{4}{|c|}{UF-IQF}\\ \hline
&{$d_1$}&{$d_2$}&{$d_3$}& &{$d_1$}&{$d_2$}&{$d_3$} \\ \hline
$q_1$&20&0&0&$q_1$&4.4&0&0 \\ \hline
$q_2$&10&10&0&$q_2$&2.2&16.1&0 \\ \hline
$q_3$&10&0&2&$q_3$&2.2&0&1.84 \\ \hline
$q_4$&5&0&10&$q_4$&1.1&0&9.2 \\ \hline
\hline
\multicolumn{4}{|c|}{UFW-IQF}&\multicolumn{4}{|c|}{UFW-IUF}\\ \hline
{$q(d_j)$}&{\em 4}&{\em 1}&{\em 2}&{$u(d_j)$}&{\em 3}&{\em 2}&{\em 2} \\ \hline
{$IQF$}&{\em 0.22}&{\em 1.61}&{\em 0.92}&{$IUF$}&{\em 0.29}&{\em 0.69}&{\em 0.69} \\ \hline
$q_1$&0.17&0&0&$q_1$&0.22&0&0 \\ \hline
$q_2$&0.14&1.04&0&$q_2$&0.19&0.45&0 \\ \hline
$q_3$&0.16&0&0.42&$q_3$&0.21&0&0.32 \\ \hline
$q_4$&0.13&0&0.64&$q_4$&0.16&0&0.48 \\ \hline
\end{tabular}
\label{tab:query-represent}
\end{table}

\subsection{Global Consistency Model}
As mentioned in section~\ref{sec:gcongraph}, the global consistency model focuses on the manner of utilizing user frequency that is consistent to the global nature of the URL. Since the global nature of each URL represents its influence over a given query on user click graph, it is considered to be more informative that the local user frequency when representing a query. Following the rules proposed in section~\ref{sec:global-consistency}, the representation for the query with global consistency can be formalized as follow:
\begin{equation}v_{ij} = \frac{g(d_j)}{log(e + \frac{\sum_{j\in D}uf_{ij}}{uf_{ij}})}\end{equation}
The global consistency model complies with the three rules we have defined. In particular, with respect to Rule~\ref{item:rule1}, if $g(d_j) > g(d_k)$, in order to achieve $v_{ij} > v_{ik}$ (global consistency) while utilizing user frequency, we add a base {\em e} inside the log function to alleviate impact of the difference between $uf_{ij}$ and $uf_{ik}$, which reduces the possibility of breaching global consistency to a large extent. Obviously, global consistency model complies with Rule~\ref{item:rule2} and~\ref{item:rule3}.

If inverse query frequency is applied to specify the global weight of the URL, it forms one implementation for global consistency model in (\ref{eql:UFW-IQF}) - {\em user frequency weighted inverse query frequency} (UFW-IQF):
\begin{equation}v_{ij} = \frac{log(\frac{|Q|}{q(d_j)})}{log(e + \frac{\sum_{j\in D}uf_{ij}}{uf_{ij}})}
\label{eql:UFW-IQF}
\end{equation}
Besides, we have also discussed IUF as the global weight of URL in section~\ref{sec:gcongraph}, leading to another implementation in (\ref{eql:UFW-IUF}) - {\em user frequency weighted inverse URL frequency} (UFW-IUF):
\begin{equation}v_{ij} = \frac{log(\frac{|U|}{u(d_j)})}{log(e + \frac{\sum_{j\in D}uf_{ij}}{uf_{ij}})}
\label{eql:UFW-IUF}
\end{equation}
We use keywords ``user frequency weighted'' to name the solution in order to emphasize the importance of global nature on user click graph, and user frequency is utilized in tune with the global nature of URLs. The query representation for Figure~\ref{fig:fig1} using different models is presented in Table~\ref{tab:query-represent}, in which the edge value is consistent to the global weight of URL within UFW-IQF and UFW-IUF. The next two sections will propose the data collection and experimental result about the performance of these models under empirical query similarities.

\section{Data Collection}\label{sec:datacollection}
The experimental dataset we used is the AOL search engine log~\cite{Pass:Picture}. The log data comprises two million search keywords for over 650,000 users over a 3-month period, in the form of UserID, Query, Time, Rank, and ClickURL, see Table~\ref{tab:aol}. In particular, there are 19,046,152 effective user clicks, and then we remove stop words and punctuation from the queries which are converted to lowercase letters. Furthermore, the queries which have less than 4 user-clicks~\cite{Wang:Organize} have been eliminated in order to reduce noise. Finally, we obtain 675,170 distinct queries with 6.84 average clicks per query, and 871,676 distinct URLs with 5.30 average clicks per URL. Noted that we only consider user frequency in which the same query-URL pair by the same user only count once.

\begin{figure}
\begin{center}
\begin{tabular}{cc}
\resizebox{40mm}{!}{\includegraphics{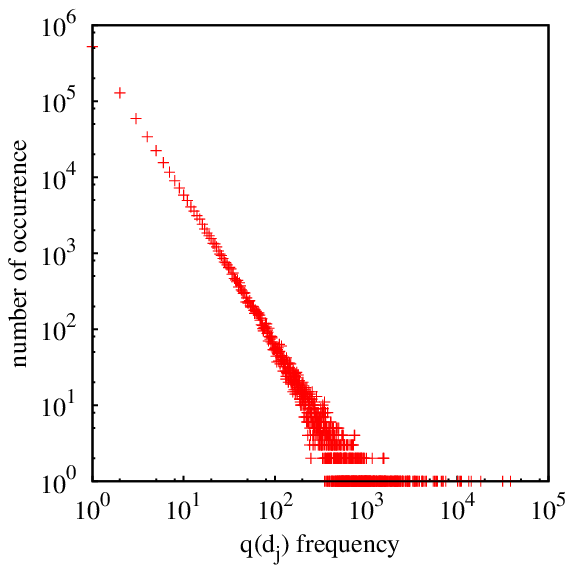}} &
\resizebox{40mm}{!}{\includegraphics{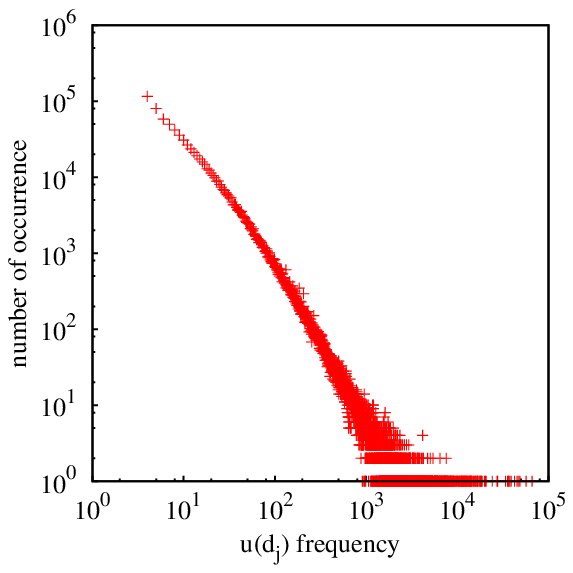}} \\
(a) & (b) \\
\end{tabular}
\caption{Distribution of $q(d_j)$ and $u(d_j)$.}
\label{fig:freq}
\end{center}
\end{figure}

\begin{table}
\centering
\caption{AOL log example}
\renewcommand{\arraystretch}{1.5}
\small
\begin{tabular}{cp{10mm}p{17mm}{c}cp{20mm}} \hline
URL&Query&Time&Rank&ClickURL\\ \hline
217&lottery&2006-03-27 16:34:59&1&calottery.com\\
217&ask.com&2006-03-31 14:31:10&1&ask.com\\
1326&konig wheels&2006-04-18 13:29:52&2&konigwheels.com\\
\hline
\end{tabular}
\label{tab:aol}
\end{table}

\begin{table}
\centering
\caption{Coefficients of power low distribution}
\renewcommand{\arraystretch}{1.5}
\begin{tabular}{ccc} \hline
$y=Ax^{-B}$&A&B\\ \hline
$q(d_j)$&31395&1.45\\ \hline
$u(d_j)$&33575&1.56\\ \hline
\end{tabular}
\label{tab:coefficient}
\end{table}

After data cleaning, we observe the power law distribution~\cite{Baeza-Yates:Semantic,Xue:Optimizing} of the click data. Figure~\ref{fig:freq}(a) is the distribution of query (user) frequency with the number of URLs versus $q(d_j)$, and the distribution of $u(d_j)$ we have discussed in section~\ref{sec:global-nature} is shown in Figure~\ref{fig:freq}(b), with the number of URLs versus $u(d_j)$. Furthermore, we compute the coefficients of the two power law distributions using regression analysis. As listed in Table~\ref{tab:coefficient}, the value of {\em B} from $u(d_j)$ frequency is larger than the one from $q(d_j)$ frequency, which implies that with the same scale of frequency variation ($\Delta x$), $u(d_j)$ will have a smaller variation in the corresponding number of URLs ($\Delta y$) hence demonstrates a smoothing effect over $q(d_j)$. Because with respect to a certain URL $d_j$, $u(d_j) < q(d_j)$ due to similar queries associated with $d_j$ (query-clicks concentrate on a smaller number of URLs) while $u(d_j) > q(d_j)$ in the case of diverse queries (query-clicks diverge to larger number of URLs). Therefore as for capturing the global weight of URL, inverse URL frequency which makes use of $u(d_j)$ is superior to inverse query frequency, which relies on $q(d_j)$.


\section{Experiments}\label{sec:exp}

\subsection{Query Similarities}
The query representations for all the models are normalized from the edge value $v_{ij}$ (query $q_i$ to URL $d_j$) to form the transition probability~\cite{Poblete:Unified}:
\begin{equation}p(d_j|q_i) = \frac{v_{ij}}{\sum_{j\in D}v_{ij}}\end{equation}
\begin{equation}p(q_i|d_j) = \frac{v_{ij}}{\sum_{i\in Q}v_{ij}}\end{equation}
Hence we can compute the transition probability matrices $P_{q2d}$ $(R^{M\times N})$ for query-to-document and $P_{d2q}$ $(R^{N\times M})$ for document-to-query. A query can be represented as $q_i$ = [$p_{q2d}(i,1)$, $p_{q2d}(i,2)$,..., $p_{q2d}(i,N)$] while a document can be expressed as $d_j$ = [$p_{d2q}(j,1)$, $p_{d2q}(j,2)$,..., $p_{d2q}(j,M)$].

We consider three methods for the query similarity experiment - cosine, jaccard and personalized PageRank to test the performance of different query representation models. The cosine similarity is the most popular method to measure the similarity between two data point:
\begin{equation}cosine(q_i,q_j) = \frac{q_i \cdot q_j}{|q_i||q_j|}\end{equation}
the jaccard coefficient defines the similarity between two data points as the intersection divided by the union:
\begin{equation}jaccard(q_i,q_j) = \frac{\sum_{n\in N}P_{q2d}(i,n)\cap P_{q2d}(j,n)}{\sum_{n\in N}P_{q2d}(i,n)\cup P_{q2d}(j,n)}\end{equation}

Random walk is considered as a stochastic process on the general graph, with the transition probability matrix $P_{q2d}$ and $P_{d2q}$ on the query-URL bipartite graph, the similarity between query $q_i$ and query $q_j$ can be defined as the transition probability from $q_i$ to $q_j$:
\begin{equation}p(q_j|q_i) = \sum_{k\in D}p(d_k|q_i)p(q_j|d_k)\end{equation}
The above random walk transition probability can be represented as an entry of transition probability matrix from query to query $P_{q2q}$ $(R^{M\times M})$, which is the 2-step transition where $p(q_j|q_i)$ = $P_{q2q}(i,j)$, and the number of walking steps for query to query on a bipartite graph is {\em 2s} steps (s = 1, 2,...)~\cite{Gao:Smoothing}. The personalized PageRank~\cite{Haveliwala:Personalizing} is a query sensitive extension of PageRank by smoothing the Markov chain~\cite{Szummer:Markov} with a query specific jumping probability vector instead of a uniform vector, to achieve the query-dependent ranking:
\begin{equation}R^{n+1}_j = (1-\alpha)R^{n}_j + \alpha\sum_{i}P(q_j|q_i)R^{n}_i\end{equation}
n represents the number of steps, when $n = 0$ $R_0$ becomes the personalized vector in which $R^0_j$ = 1 if vertex {\em j} is the initial query otherwise $R^0_j$ = 0. The value of jumping constant $\alpha$ is between 0 and 1, which scales the rate of propagation. The performance of personalized PageRank with different values of jumping constant is presented in section~\ref{sec:performance}.

\subsection{Evaluation Metric}
We employ the same method as used in~\cite{Baeza-Yates:Semantic,Deng:Entropy-biased}  for automatic evaluation, using Google Directory to represent each query. Specifically, we extract the categories for each query from Google, and hence each query will be represented by one or more directory paths. For example, the query `haiti' may be denoted by the directory path `Regional > Caribbean > Haiti > Guides-and-Directories' and the query `haiti news' may be denoted by `Regional > Caribbean > Haiti > News-and-Media'. The similarity between two directory path $D_i$, $D_j$ is defined as: $sim(D_i, D_j)$ = $|P(D_i, D_j)|$/$max(|D_i|,|D_j|)$, where $P(D_i, D_j)$ is the longest common prefix between the two paths. Therefore the similarity between `haiti' and `haiti new', with the longest common prefix 3 and maximum length 4, is $3/4$, while the similarity between `haiti', and `haiti history' of which the directory path is `Society > History > By-Region > Caribbean > Haiti', is $2/5$.

We randomly select 500 queries, and retrieve the top 10 similar queries for each one and the corresponding top 5 Google directories for each query terms, the similarity of the two groups of directory paths (two queries) is chosen to be the most similar categories between the two. Given a query $q$ and the retrieved result $q_r$, the evaluation of rank n precision for the top n query results is defined as follow:
\begin{equation}p@n = \frac{\sum_{r=1}^n{sim(q,q_r)}}{n}\end{equation}
where $sim(q, q_r)$ is the similarity between $q$ and $q_r$. Besides, we also evaluate the specificity of the results in terms of query length, for example `metropole haiti news' ($L = 3$) is more specific than `haiti news' ($L = 2$). The specificity of the top n query results is defined as follow:
\begin{equation}L@n = \frac{\sum_{r=1}^n{L(q_r)}}{n}\end{equation}
where $L(q_r)$ is the length of $q_r$. For cosine and jaccard, we have computed the average precision of query similarities at different ranks, and the average length of results at rank 10. The average precision for personalized PageRank is presented at different steps with different jumping constants.

\begin{figure}
\begin{center}
\includegraphics{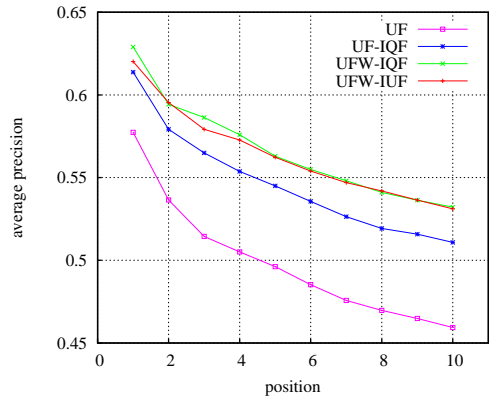}
\caption{Average precision at different positions with cosine similarity.}
\label{fig:cosine}
\end{center}
\end{figure}
\begin{figure}
\begin{center}
\includegraphics{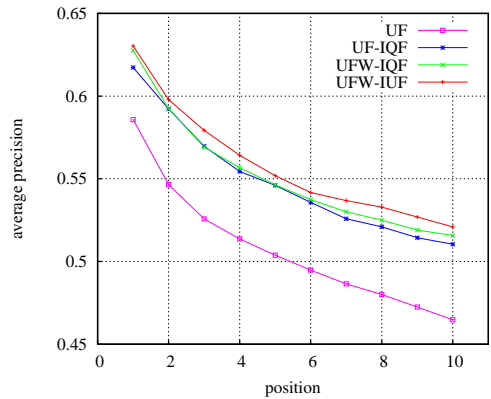}
\caption{Average precision at different positions with jaccard coefficient.}
\label{fig:jaccard}
\end{center}
\end{figure}

\subsection{Performance Analysis}\label{sec:performance}
We have tested the performance of four models: UF, UF-IQF, UFW-IQF and UFW-IUF, and report their average precision from $P@1$ to $P@10$ and average result length $L@10$ under empirical query similarities. The results show our global consistency model (UFW-IQF or UFW-IUF) not only achieves the best average precision with popular query similarities at different positions, as can be seen from Figure~\ref{fig:cosine} and Figure~\ref{fig:jaccard}, but also boosts long tail results related to the query compared to other models, indicated by $L@10$ in Table~\ref{tab:performance}. The precision boost from global consistency model on cosine similarity is very clear, with $4.17\%$ improvement over entropy-biased model, and $15.9\%$ over user frequency at rank 10. The improvement on jaccard similarity is relatively small, in which UFW-IUF improves $1.01\%$ over UF-IQF, and $12.1\%$ over UF. As to inverse query frequency (IQF) and inverse URL frequency (IUF), which are two different schemes to capture the global weight URL, IUF performs better than IQF in most cases, though their precisions on cosine similarity are very close.

\begin{table}
\centering
\caption{Comparison of average precision P@1, P@10, and average query length L@10}
\renewcommand{\arraystretch}{1.2}
\small{
\begin{tabular}{|c|c|c|c|c|c|} \hline
\multicolumn{2}{|c|}{method}&UF&UF-IQF&UFW-IQF&UFW-IUF\\ \hline
\multirow{3}{*}{\begin{sideways}cosine\end{sideways}}
&P@1&0.5774&0.6138&{\bf 0.6290}&0.6202\\
&P@10&0.4594&0.5109&{\bf 0.5322}&0.5311\\
&L@10&2.4878&2.6342&2.8782&{\bf 2.8918}\\ \hline
\multirow{3}{*}{\begin{sideways}jaccard\end{sideways}}
&P@1&0.5858&0.6173&0.6277&{\bf 0.6303}\\
&P@10&0.4647&0.5104&0.5157&{\bf 0.5210}\\
&L@10&2.7632&2.8368&2.8482&{\bf 2.8552}\\ \hline
\end{tabular}
}
\label{tab:performance}
\end{table}

\begin{table}
\centering
\caption{Examples of query result of different models with cosine similarity}
\renewcommand{\arraystretch}{1.3}
\small{
\begin{tabular}{|c|l|l|} \hline
&query = `haiti'&query = `shiny cowbird'\\ \hline \hline
\multirow{5}{*}{\begin{sideways}UF\end{sideways}}
&www haiti.com&tarpon\\
&djibouti&wikipedia\\
&cia&jack dunphy\\
&cameroon&satire\\ \hline
\multirow{5}{*}{\begin{sideways}UF-IQF\end{sideways}}
&www haiti.com&yellow breasted bird florida\\
&cia&florida fish game\\
&djibouti&fishing license\\
&cia world factbook&yellow finches map\\ \hline
\multirow{5}{*}{\begin{sideways}UFW-IQF\end{sideways}}
&www haiti.com&yellow breasted bird florida\\
&haiti news&yellow finches map\\
&madagascar country&graylag geese breeder\\
&news port au prince haiti&red throated bird\\ \hline
\end{tabular}
}
\label{tab:result-example}
\end{table}

In Table~\ref{tab:result-example} we have listed some query similarity examples obtained from different models. For instance when query = `shiny cowbird', compared with the UF model, the entropy-biased model (UF-IQF) rejects some irrelevant queries like `wikipedia' and `satire' by downplaying the corresponding edges `shiny cowbird - www.wikipedia.org' and `satire - www.wikipedia.org', in which the IQF value of the URL `www.wikipedia.org' is small since it is highly clicked. The global consistency model achieves the same thing while providing some more relevant long tail queries such as `red throated bird'. Moreover, take the query `haiti' as another example, although the IQF value of the URL `www.cia.gov/ .../the-world-factbook' is small, the updated edge value `haiti - www.cia.gov/.../the-world-factbook' and `djibouti - www.c- ia.gov/.../the-world-factbook' are still very large in the entropy biased model due to high user frequency. However, in the global consistency model, those updated edge values will be consistent to the IQF value of the URL, hence it not only creates the opportunities for other long tail queries with low user frequency, but also guarantees the relevance between the result and the query by global consistency.
\begin{figure}[t]
\begin{center}
\begin{tabular}{c}
\includegraphics{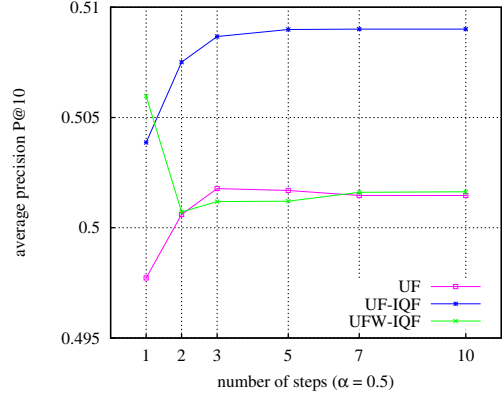}\\
(a)\\
\includegraphics{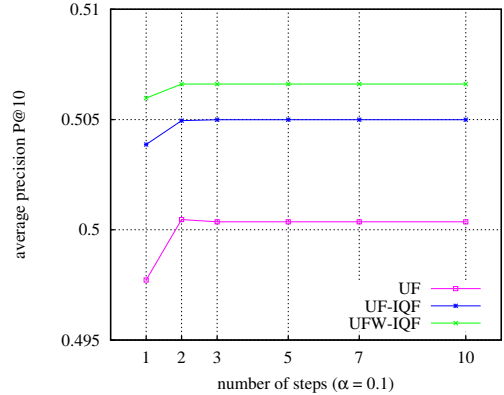}\\
(b)\\
\end{tabular}
\caption{Average precision of personalized PageRank at rank 10.}
\label{fig:pagerank}
\end{center}
\end{figure}

\begin{table}
\centering
\caption{Comparison of average length L@10 of query results with personalized PageRank}
\begin{tabular}{|c|c|c|c|} \hline
jumping constant&UF&UF-IQF&UFW-IQF\\ \hline
$\alpha = 0.5$&2.27&2.30&{\bf 2.77}\\ \hline
$\alpha = 0.1$&2.30&2.33&{\bf 2.80}\\ \hline
\end{tabular}
\label{tab:length-pagerank}
\end{table}

The average precisions of different models on personalized PageRank are very close compared with the performance on cosine or jaccard similarities, instead it is the behavior of each model that draws our attention. With more steps walked on the click graph, the result approaches to the stationary distribution. When the jumping constant $\alpha = 0.5$, as indicated in Figure~\ref{fig:pagerank}(a), the global consistency model (UFW-IQF) experiences a significant drop of average precision at step 2, while the behavior of other models follows the convention with better precision at step 2. If the jumping constant $\alpha = 0.1$ which implies an extremely low rate of propagation on the graph, and the first step becomes the dominant factor since the result varies little in subsequent steps. As shown in Figure~\ref{fig:pagerank}(b), the average precision of each model is determined by the first step. The personalized PageRank reveals the nature of those different models: for the global consistency model, most of the relevant queries have already been retrieved during the first step, while subsequent steps are needed for other models to retrieve the most relevant results with a reasonable jumping constant. In other words, the conventional models need a proper propagation rate to walk on the click graph in order to achieve the best performance, while within global consistency model, the best results are closely clung to the initial preferred vertex without much propagation. In Table~\ref{tab:length-pagerank}, we also listed the average length of results at rank 10 with personalized PageRank, we observe that on one hand, a smaller jumping constant (lower propagation rate) favors long tail results within each model, since the propagation on the graph is heading to the queries with larger transition probability which tend to be the popular short tail queries; on the other hand, the global consistency model (UFW-IQF) demonstrates the superiority of boosting long tail queries over other models despite of different jumping constants.

\section{Conclusion}\label{sec:conclusion}
In this paper we propose a novel model for query representation on user click  graph, which is based on the observation that for a certain query, the global nature of URLs is more informative than local user frequency. The global consistency model identifies the inverse query frequency from previous work as a global property of the URL, based on which we suggests a more effective scheme called inverse URL frequency which further considers the similarities among queries for global nature capturing. Besides, we formalize the framework of utilizing user frequency in tune with the global nature of the URL. The global consistency model consistently demonstrates better performance over current models for query representation under popular query similarities, in terms of better precision and long tail result boost. Hence many query log analysis tasks can benefit from this query representation model.

\section*{Acknowledgement}\label{sec:ack}
This work is supported by the National Natural Science Foundation of China
(Grant Nos. 61103185, 61073118 and 61003247), the Start-up Foundation of Nanjing Normal University (Grant No. 2011119XGQ0072), Natural Science Foundation of the Higher Education Institutions of Jiangsu Province, China (Grant No. 11KJB520009), the 9th Six Talents Peak Project of Jiangsu Province (Grant No. DZXX-043), and
Major Program of National Natural Science Foundation of Jiangsu Province (Grant No. BK2011005).


\end{document}